\documentclass[nofootinbib,notitlepage,superscriptaddress,10pt,aps,pra,twocolumn]{revtex4-1}

\usepackage[utf8x]{inputenc}
\usepackage{amsmath}
\usepackage{amssymb}
\usepackage{graphicx}
\usepackage[normalem]{ulem}
\usepackage{epsf,epsfig}
\usepackage{psfrag}

\begin{document}

\title{Spacetime-noncommutativity regime of Loop Quantum Gravity}

\author{Giovanni AMELINO-CAMELIA}
\affiliation{Dipartimento di Fisica, Universit\`a di Roma ``La Sapienza", P.le A. Moro 2, 00185 Roma, Italy}
\affiliation{INFN, Sez.~Roma1, P.le A. Moro 2, 00185 Roma, Italy}

\author{Malú Maira DA SILVA}
\affiliation{Dipartimento di Fisica, Universit\`a di Roma ``La Sapienza", P.le A. Moro 2, 00185 Roma, Italy}
\affiliation{INFN, Sez.~Roma1, P.le A. Moro 2, 00185 Roma, Italy}

\author{Michele RONCO}
\affiliation{Dipartimento di Fisica, Universit\`a di Roma ``La Sapienza", P.le A. Moro 2, 00185 Roma, Italy}
\affiliation{INFN, Sez.~Roma1, P.le A. Moro 2, 00185 Roma, Italy}

\author{ Lorenzo CESARINI}
\affiliation{Dipartimento di Fisica, Universit\`a di Roma ``La Sapienza", P.le A. Moro 2, 00185 Roma, Italy}

\author{ Orchidea Maria LECIAN}
\affiliation{Dipartimento di Fisica, Universit\`a di Roma ``La Sapienza", P.le A. Moro 2, 00185 Roma, Italy}

\begin{abstract}
A recent study by Bojowald and Paily ~\cite{bojopaily} provided a path toward the identification of an effective quantum-spacetime picture
of Loop Quantum Gravity, applicable in the ``Minkowski regime", the regime where the large-scale (coarse-grained) spacetime metric is flat.
A pivotal role in the analysis is played by  Loop-Quantum-Gravity-based modifications to
the hypersurface deformation algebra, which leave a trace in  the Minkowski regime.
We here show that the symmetry-algebra results reported by
Bojowald and Paily are consistent with a description of spacetime in the Minkowski regime given in terms
of the $\kappa$-Minkowski noncommutative spacetime, whose relevance for the study of the quantum-gravity problem had already been
 proposed for independent reasons.
\end{abstract}
\maketitle

\section{INTRODUCTION}
Over the last decade the quantum-gravity literature has been increasingly polarizing into a top-down approach and a bottom-up approach.
The top-down approach attempts to provide models that could potentially solve at once all aspects of the quantum-gravity problem,
but typically involves formalisms of very high complexity, rather unmanageable for obtaining physical intuition about observable (and potentially testable) features.
The bottom-up approach relies on relatively simpler models, suitable for describing only a small subset of the departures from standard
 physics that the quantum-gravity realm is expected to host, but has the advantage of producing better opportunities
 for experimental testing ~\cite{gacLRR}.
 A good synergy between the two approaches would be desirable: from the top we could obtain guidance
 on which are the most significant structures to be taken into account in more humble formalizations, and from the bottom
 we could develop insight on how to handle those structures, hopefully also in terms of experimental tests.
 Unfortunately very little
 has been accomplished so far in the spirit of such a synergy, but
 we here offer a contribution toward establishing a link between Loop Quantum Gravity (LQG) ~\cite{RovelliLRR, AshLew, Thiem}, one of the most studied approaches aiming for a full
 solution of the quantum-gravity problem, and approaches focusing on the assumption of noncommutativity of coordinates in the Minkowski
 regime of quantum gravity ~\cite{gacMajid,majid,kowNow,girellivOriti}.

 Both the LQG approach and the spacetime-noncommutativity approach involve the possibility that the geometry of spacetime
 might be quantized in the quantum-gravity realm. In LQG one in principle obtains a very powerful picture of this quantum geometry,
 applicable to all regimes of quantum gravity, but the complexity of the formalism is such that {\it de facto}
there is no physical regime of quantum gravity for which we are presently able to use LQG for an intuitive (intelligible) characterization
of the novel physical properties that would result from the quantum-geometric properties. Spacetime noncommutativity takes the more humble
approach of postulating one or another form of noncommutativity of spacetime coordinates, hoping
that it might be applicable in the Minkowski regime, but has the advantage of leading to several rather intuitive findings
about the physical implications of these assumptions, some of which attracted even some interest in phenomenology ~\cite{gacLRR,MAJIDgrb,bala,gacEllisMavNanSak,gacPiran}.
A link between LQG and spacetime noncommutativity is solidly established for dimensionally-reduced 3D quantum gravity ~\cite{FREIDEL,ORITI,scaef,gacfreidkowsmol},
but it remains so far unclear whether a generalization of those results is applicable to the 4D case of real physical interest.

We here take what might be a significant step toward establishing a link between (4D) LQG and spacetime noncommutativity.
The starting point for our analysis is provided by a recent publication by Bojowald and Paily ~\cite{bojopaily}, which contemplated
some LQG-based modifications to
the hypersurface deformation algebra. Interestingly, Bojowald and Paily found that these modifications
of the hypersurface deformation algebra leave a trace in the Minkowski regime, characterized by a suitable modification
of the Poincar\'e algebra, and correctly concluded that a quantum-spacetime dual to that deformed Poincar\'e algebra
should then give the quantum-geometry description of LQG in the Minkowski regime.
Indeed deformed Poincar\'e algebras are characteristic of the structure of DSR-relativistic theories,
first introduced in Ref. ~\cite{gacIntJ} (also see the follow-up studies in Refs. ~\cite{magsmolin, kowNow,gacModP,gacPhyLett}),
and for such theories one expects in general that the duality between Minkowski spacetime and the classical Poincar\'e algebra
be preserved in the form of a duality between a suitably deformed  Poincar\'e algebra and a ``quantum Minkowski spacetime".
Bojowald and Paily also contemplated the possibility that this quantum-spacetime picture be given in terms
of the much-studied ``$\kappa$-Minkowski noncommutative spacetime" ~\cite{majRue,lukRueg}, but explored this possibility only preliminarily by relying
on a specific  {\it ansatz} for the representations of the action of the relevant
deformed Poincar\'e algebra on $\kappa$-Minkowski coordinates. This preliminary exploration gave negative results (incompatibility
between the {\it ansatz} for the representations and some properties of $\kappa$-Minkowski coordinates), leading
Bojowald and Paily to tentatively conclude that $\kappa$-Minkowski might not provide the needed quantum-spacetime picture.

We here expose some limitations of the {\it ansatz} adopted by Bojowald and Paily. By considering a more general class
of possible representations of the action of the relevant
deformed Poincar\'e algebra on $\kappa$-Minkowski coordinates we establish the compatibility between
$\kappa$-Minkowski and the findings reported by Bojowald and Paily for the LQG-deformed hypersurface deformation algebra.
Our analysis also leads to the identification of the ``coproduct structure" which is to be expected in regimes characterized
by $\kappa$-Minkowski noncommutativity, relevant for the description of the action of relativistic-symmetry transformations
on products of fields. Finding evidence of this coproduct structure on the LQG side would provide final proof of
the correspondence between LQG and $\kappa$-Minkowski.

\section{DEFORMED HYPERSURFACE-DEFORMATION ALGEBRA}

The Hamiltonian formulation of general relativity allows to encode the general covariance of the theory in the algebra closed by the scalar ($H[N]$) and vector ($D[N^{i}]$) constraints, the so-called Hypersurface-Deformation Algebra (HDA) ~\cite{Dirac, ADM, CorichiReyes}:
\begin{equation}\label{hda}
\begin{split}
\{D[M^{k}],D[N^{j}]\}=D[\mathcal{L}_{\vec{M}}N^{k}],\\
\{D[N^{k}],H[M]\}=H[\mathcal{L}_{\vec{N}}M],\\
\{H[N],H[M]\}=D[h^{jk}(N\partial_{j} M-M\partial_{j} N)],
\end{split}
\end{equation}
where $H[N]$ and $D[N^{i}]$ depend respectively on the lapse $N$ and shift $N^{i}$ functions~\cite{ADM}.

It is natural to ask if the HDA should be deformed in quantum gravity due to the presence of quantum-geometry corrections, such as those arising in LQG ~\cite{CaitellMielcBarr, AshtLewMarMouThie, bojopail2,bojoHossKag}. Indeed, recently there has been a growing effort in studying quantum deformations of Eqs. \eqref{hda}, especially in the context of models motivated by LQG ~\cite{bojopail2,bojoHossKag,Calcagni,Barrau}.
For our purposes here, it is of particular interest the analysis reported in Ref. ~\cite{bojopaily}, which shows that a particular case of deformed HDA reduces to a Planck-scale-deformed Poincaré algebra if one takes the flat-spacetime limit. Most notably the relevant deformations of the Poincaré algebra are qualitatively of the type known to arise in the description of the relativistic symmetries of noncommutative spacetimes.

From a wider perspective this shows a possible path from ambitious quantum-gravity theories to certain phenomenological opportunities
available~\cite{gacLRR} in the ``Minkowski regime": within the full quantum-gravity theory one should be in position to study
the deformations of the HDA; then by taking the Minkowski-regime limit of such a deformed HDA one should find the corresponding
 deformation of the Poincaré algebra applicable to the Minkowski limit of the model, and finally one could obtain an effective quantum-Minkowski-spacetime picture, by duality with the relevant deformed Poincaré algebra.
 The analysis reported by Bojowald and Paily~\cite{bojopaily} provides an opportunity for a first application of this strategy of analysis,
 relevant for the LQG approach.
 The first two steps have already been accomplished in Ref.~\cite{bojopaily}, by motivating a specific deformation of the HDA and by finding
 the associated deformation of the Poincaré algebra applicable to the Minkowski limit.
 For the third step, the one providing an effective
 quantum-Minkowski-spacetime picture, the analysis reported in Ref.~\cite{bojopaily} was inconclusive for reasons already discussed
 in the previous section.
Since our objective here is to take this  important third step, we find appropriate to summarize briefly in this section some key aspects of the LQG-deformed HDA, from the perspective adopted in Ref.  ~\cite{bojopaily}.

As mentioned,  it is expected that Eqs. \eqref{hda} should receive quantum-gravity corrections~\cite{Calcagni,CaitellMielcBarr,PerezPranz,Loll},
and in particular this is expected for the LQG scenario~\cite{bojopail2, Barrau, CaitellMielcBarr, Thiemann2}. A fully deductive derivation of the deformed HDA within LQG is at present beyond our technical abilities, so different authors have relied on different approximation schemes, but all results agree on the following form for the deformed HDA~\cite{Calcagni, bojopail2, Barrau, CuttSakell}:
\begin{equation}\label{qhda}
\begin{split}
\{D[M^{a}],D[N^{a}]\}=D[\mathcal{L}_{\vec{M}}N^{a}],\\
\{D[N^{a}],H^{Q}[M]\}=H^{Q}[\mathcal{L}_{\vec{N}}M],\\
\{H^{Q}[M],H^{Q}[N]\}=D[\beta h^{ab}(M\partial_{b}N-N\partial_{b}M)],
\end{split}
\end{equation}
where $H^{Q}[N]$ denotes a deformed (``quantum") scalar constraint and $\beta$ depends on the specific corrections that are taken into account. A key challenge is to find an appropriate representation of constraints as operators on a Hilbert space, and so far no proposal has fully accomplished this task. However,  several techniques have been developed and some promising candidates for the quantum Hamiltonian operator ($H^{Q}[N]$) have been proposed~\cite{Thiem, Thiemann2, AshtLewMarMouThie, bojoBraRey, AlesciCianf}. In particular, in spherically-symmetric models~\cite{bojopail2, bojoHossKag}, which are here of interest, some of the quantum corrections, namely the local (i.e. point-like) holonomy corrections~\cite{bojopail2, bojoHossKag, bojoBraRey}, have been successfully implemented, and the corresponding quantized version of the scalar constraint can still close an algebra provided that it is properly deformed as in Eqs. \eqref{qhda}.

For spherically-symmetric analyses it is useful to write the 3-metric $h_{ij}$ in suitably adapted fashion, and for this purpose it is useful
to describe the densitized triads~\cite{Thiem,Beng} as follows:
\begin{equation}\label{triads}
\begin{split}
E=E^{a}_{i}\tau^{i}\frac{\partial}{\partial x^{a}}=E^{r}(r)\tau_{3}\sin\theta\frac{\partial}{\partial r}+\\+E^{\varphi}(r)\tau_{1}\sin\theta\frac{\partial}{\partial\theta}+E^{\varphi}(r)\tau_{2}\frac{\partial}{\partial\varphi},
\end{split}
\end{equation}
where $\tau_{j}=-\frac{1}{2}i\sigma_{j}$ represent \textit{SU}(2) generators. The
densitized triads are canonically conjugate to the extrinsic curvature components, which, in presence of spherical symmetry, are conveniently
described as follows~\cite{bojopail2, bojoHossKag}:
\begin{equation}
\begin{split}
K=K^{i}_{a}\tau_{i}dx^{a}=K_{r}(r)\tau_{3}dr+K_{\varphi}(r)\tau_{1}d\theta+\\+K_{\varphi}(r)\tau_{2}\sin\theta d\varphi.
\end{split}
\end{equation}
It has been shown ~\cite{bojopail2, bojoHossKag, bojoBraRey} that the deformation function $\beta$ depends on the angular component of the extrinsic curvature $K_{\varphi}$ as follows:
$$\beta=\cos(2\delta K_{\varphi}) \, ,$$
where $\delta$ is a parameter which can be related to the square root of the minimum eigenvalue of the area operator~\cite{Thiemann2,bojoBraRey}.
Since we would like to obtain, in the appropriate limit, a deformed Poincar\'e algebra, it is convenient to write $\beta$ in terms of symmetry generators, and for this purposes it is valuable to observe that observables of the Brown York momentum~\cite{BY},
\begin{equation}\label{bymom}
P=2\int_{\partial\Sigma} d^{2}z\upsilon_{b}(n_{a}\pi^{ab}-\overline{n}_{a}\overline{\pi}^{ab}) \, ,
\end{equation}
can be identified by extrinsic curvature components provided a suitable choice for $\delta \propto |E^{r}|^{-\frac{1}{2}}$. In Eq.\eqref{bymom}, we have that $\upsilon_{a}=\partial/\partial x^{a}$, $n_{a}$ is the conormal of the boundary of the spatial region $\Sigma$, and $\pi^{ab}$ plays the role of the gravitational momentum. From this, it is possible to establish that the radial Brown-York momentum $P_{r}$ is related to the extrinsic curvature component $K_{\varphi}$ in the following way ~\cite{bojopaily}:
\begin{equation}\label{bymomf}
P_{r}=-\frac{K_{\varphi}}{\sqrt{|E^{r}|}}.
\end{equation}
In order to obtain the Poincaré algebra from the HDA one should consider flat spatial slices of spacetime, {\it i.e.} a Euclidean three metric $h_{ij} = \delta_{ij}$, and also take a combination~\cite{ReggeTeit}
of the Killing vectors of Minkowski spacetime as lapse and shift functions:
\begin{equation}
\begin{split}
N = \Delta t + v_{k}x^{k} \\
N^{i} = \Delta x^{i}+ \phi^{j}\epsilon^{ijk}x^{k}
\end{split}
\end{equation}
It is generally expected (see Ref.~\cite{bojopaily,Mielcz} and references therein)
that this very direct connection between Poincaré algebra and HDA should still be present when quantum-gravity effects are taken
into account, and therefore if the HDA is affected by quantum-gravity modifications, then also the Poincaré algebra should be correspondingly modified. We then turn to the deformed HDA of Eq.\eqref{qhda}, and notice that, in presence of spherical symmetry,
 and taking into account Eq.\eqref{bymomf}, the deformation function $\beta$ is a function of the generator of spatial translations, {\it i.e.} $\beta = \cos(\lambda P_{r})$, where $\lambda$ is a parameter of the order of the Planck length. The net result is
 that, as a result of  Eq. \eqref{qhda}, the relevant Poincaré algebra is characterized by a deformed commutator between boost generator
 and generator of time translations:
\begin{equation}\label{bjcommut}
[B_{r}, P_{0}] = iP_{r}\cos(\lambda P_{r}) \, .
\end{equation}
Since only the Poisson bracket involving two scalar constraints  is quantum corrected (see Eqs. \eqref{qhda}), the other commutators are undeformed, {\it i.e.} $[B_{r}, P_{r}] = iP_{0}$ and $[P_{0}, P_{r}] = 0$.

\section{COMPATIBILITY WITH $\kappa$-MINKOWSKI}
Our next task is to show that the operators $B_{r}, P_{r}$ and $P_{0}$ generate the deformed-Poincar\'e-symmetry transformations which
are symmetries of the $\kappa$-Minkowski noncommutative spacetime. What is here relevant is the spherically-symmetric version
of the $\kappa$-Minkowski noncommutativity of spacetime coordinates ~\cite{majRue,lukRueg}:
\begin{equation}\label{kmink}
[X_{0}, X_{r}] = i\lambda X_{r}.
\end{equation}
As mentioned above, already  Bojowald and Paily had contemplated the possibility~\cite{bojopaily}
that $B_{r}, P_{r}$ and $P_{0}$ might generate the deformed-Poincar\'e-symmetry transformations which
are symmetries of the $\kappa$-Minkowski noncommutative spacetime. They however somehow assumed
the following representations:
\begin{equation}
\label{bojochoice}
\begin{split}
B_{r} = x_{r}p_{0}-\cos(\lambda p_{r}) x_{0}p_{r} \\
P_{r} = p_{r} \\
P_{0} = p_{0} \, ,
\end{split}
\end{equation}
in terms of standard operators such that $[x_{r}, p_{r}] = i$, $[x_{0}, p_{0}] = -i$,
$[x_{r}, p_{0}] =0$, $[x_{0}, p_{r}] =0$, $[x_{0}, x_{r}] =0$, $[p_{0}, p_{r}] =0$,
and they correctly found (also relying on results which had been previously
reported in Ref.\cite{KovMeljPachol}) that this representation is incompatible with
 the structure of $\kappa$-Minkowski spacetime.

We here notice that the representation (\ref{bojochoice}) adopted by Bojowald and Paily is only one of many possibilities
that can be tried. Experience working with spacetime noncommutativity teaches that it is actually rather hard to guess correctly such representations. A constructive approach
is usually appropriate, and we therefore seek a suitable
representation of $B_{r}, P_{r}$ and $P_{0}$ within a rather general {\it ansatz}:
\begin{equation}\label{choice2}
\begin{split}
B_{r} = F(p_{0},p_{r})X_{r}p_{0}-G(p_{0},p_{r})X_{0}p_{r} \, ,\\
P_{r} = Z(p_{r}) \, ,\\ P_{0} = p_{0} \, ,
\end{split}
\end{equation}
where $F(p_{0},p_{r})$, $G(p_{0},p_{r})$ and $Z(p_{r})$ are functions of the translation generators to be determined
by enforcing compatibility with the deformed algebra \eqref{bjcommut}. Our {\it ansatz} involves the $\kappa$-Minkowski noncommutative
coordinates $X_r , X_0$, but this is only for convenience: it is well known (see, {\it e.g.}, Ref.\cite{KovMeljPachol})
that one can represent
the coordinates $X_r , X_0$ in terms of the $x_{r}, x_{0}, p_{r}), p_{0}$ used by Bojowald and Paily
(with $X_r = x_{r}$ and $X_0 = x_{0}-\lambda x_{r}p_{r}$), so the class of representations covered by our {\it ansatz} is qualitatively
of the same type as the one of the representation considered by Bojowald and Paily. The real difference resides in the
structure of the representations and the fact that we will seek a suitable representation by determining
the functions $F(p_{0},p_{r})$, $G(p_{0},p_{r})$ and $Z(p_{r})$, rather than try to guess.

Explicitly our objective is to find choices of $F(p_{0},p_{r})$, $G(p_{0},p_{r})$ and $Z(p_{r})$ such that (\ref{bjcommut})
is satisfied, with $[B_{r}, P_{r}] = iP_{0}$ and $[P_{0}, P_{r}] = 0$. We came to notice that
this is assured if  $Z(p_{r})$ is a solution of the equation
\begin{equation}\label{condition}
\lambda Z(p_{r})\sin(\lambda Z(p_{r}))+\cos(\lambda Z(p_{r}))=\frac{\lambda^{2}{p_{r}}^{2}}{2}+1.
\end{equation}
and then  $F(p_{0},p_{r})$ and $G(p_{0},p_{r})$ are given in terms of such a solution for $Z(p_{r})$
through the following equations:
\begin{equation}\label{ansatz}
\begin{split}
G(p_{r})=\frac{Z(p_{r})\cos(\lambda Z(p_{r}))}{p_{r}},\\
F(p_{0},p_{r})=G(p_{r})e^{\lambda p_{0}}=\frac{Z(p_{r})\cos(\lambda Z(p_{r}))e^{\lambda p_{0}}}{p_{r}},
\end{split}
\end{equation}
So we have reduced the problem of finding representations on $\kappa$-Minkowski
of the Bojowald-Paily deformed Poincar\'e algebra to the problem of finding solutions
to equation \eqref{condition}.
Of course we must also enforce that such solutions $Z(p_{r})$ satisfy the limiting condition $\lim_{\lambda\to 0}\frac{Z(p_{r})}{p_{r}} = 1$,
since the undeformed representation of Poincaré generators must be recovered when the noncommutativity is turned off.

We were unable to find an explicit all-order expression for such a solution $Z(p_{r})$,
 but we find that its perturbative derivation (as a series of powers of $\lambda$)
  is always possible and straightforward up to the desired perturbative order.
   In particular, to quartic order in the parameter $\lambda$ the needed solution $Z(p_{r})$ takes the form:
\begin{equation}\label{perturbative}
Z(p_{r})=p_{r}+\frac{1}{8}\lambda^{2}{p_{r}}^{3}+\frac{55}{1152}\lambda^{4}{p_{r}}^{5}.
\end{equation}
Notice that on the basis of remarks given above evidently $p_r$ acts on
$\kappa$-Minkowski noncommutative
coordinates as follows:
\begin{equation}
[p_{r}, X_0] = i\lambda p_{r}, \quad [p_{r}, X_r] = -i
\end{equation}
while for what concerns $p_0$ one has
\begin{equation}
[p_{0}, X_0] = i, \quad [p_{0}, X_r] = 0
\end{equation}
Equipped with this final specification one can easily check explicitly that (as ensured automatically by our constructive procedure)
the representation here obtained up to quartic order in $\lambda$ for the generators $B_{r}, P_{r}$ and $P_{0}$
satisfies all the Jacobi identities involving these generators and $\kappa$-Minkowski coordinates. For example one has that:
\begin{eqnarray*}
&&[[B_{r}, X_r],X_0]+[[X_0, B_{r}],X_r]+[[X_r, X_0],B_{r}]  =\\&&-i[\frac{(Z'\cos(\lambda Z)-\lambda ZZ'\sin(\lambda Z))p_{r}-Z\cos(\lambda Z)}{p^{2}_{r}}x_{r}p_{0}, X_0]+\\&&+i[\frac{(Z'\cos(\lambda Z)-\lambda ZZ'\sin(\lambda Z))p_{r}-Z\cos(\lambda Z)}{p^{2}_{r}} x_{0}p_{r}, X_0]+\\&&-[i\frac{Z\cos(\lambda Z)}{p_{r}}x_{r}-\lambda [B_{r}, X_r]p_{r}-i\lambda\frac{Z\cos(\lambda Z)}{p_{r}}x_{r}p_{0}, X_r]+\\&&+[ i\frac{Z\cos(\lambda Z)}{p_{r}}x_{0}, X_0]+i\lambda[B_{r}, X_r] =  0
\end{eqnarray*}
where $Z' = \frac{d Z(p_{r})}{dp_{r}}$.

\section{COPRODUCTS AND CHOICE OF ``BASIS"}
The results we reported in the previous section provide strong encouragement for the possibility that
the quantum-Minkowski spacetime emerging from the Bojowald-Paily analysis is the $\kappa$-Minkowski noncommutative spacetime.
Our next objective is to discuss some implications of these findings.

The first point we make is that in order to have a description of the symmetries of a noncommutative spacetime
one should specify not only the commutator of symmetry generators but also their coproducts \cite{majRue,lukRueg}.
From the representations of $B_{r}, P_{r}$ and $P_{0}$ we derived in the previous section
one easily finds (with standard steps of derivation which have been discussed in several publications
such as Refs.\cite{majRue,lukRueg,COPRODREF}) that these coproducts
are given by:
\begin{eqnarray*}
&&\Delta B_{r}=B_{r}\otimes 1+1\otimes B_{r}-\lambda P_{0}\otimes B_{r}+\frac{1}{8}\lambda^{2}{P_{r}}^{2}\otimes B_{r}\\&& +\frac{1}{2}\lambda^{2}{P_{0}}^{2}\otimes B_{r}-\frac{3}{8}\lambda^{2}B_{r}\otimes{P_{r}}^{2}-\frac{3}{4}\lambda^{2}P_{r}B_{r}\otimes P_{r}\\&&-\frac{3}{4}\lambda^{2}P_{r}\otimes P_{r}B_{r}-\frac{5}{8}\lambda^{3}P_{0}{P_{r}}^{2}\otimes B_{r}+\frac{3}{4}\lambda^{3}P_{0}P_{r}\otimes P_{r}B_{r}\\&&-\frac{3}{4}\lambda^{3}{P_{r}}^{2}B_{r}\otimes P_{0}-\frac{3}{4}\lambda^{3}{P_{r}}^{2}\otimes P_{0}B_{r}-\frac{3}{4}\lambda^{3}P_{r}B_{r}\otimes P_{0}P_{r}\\&&-\frac{3}{4}\lambda^{3}P_{r}\otimes P_{0}P_{r}B_{r}+\frac{67}{1152}\lambda^{4}P_{r}^{4}\otimes B_{r}+\frac{15}{64}\lambda^{4}{P_{r}}^{2}\otimes{P_{r}}^{2}B_{r}\\&&-\frac{1}{8}\lambda^{4}{P_{0}}^{4}\otimes B_{r}+\frac{9}{16}\lambda^{4}{P_{0}}^{2}{P_{r}}^{2}\otimes B_{r}+\frac{15}{64}\lambda^{4}{P_{r}}^{2}B_{r}\otimes {P_{r}}^{2}\\&&-\frac{167}{288}\lambda^{4}{P_{r}}^{3}B_{r}\otimes P_{r}-\frac{59}{288}\lambda^{4}P_{r}\otimes{P_{r}}^{3}B_{r}\\&&-\frac{97}{144}\lambda^{4}{P_{r}}^{3}\otimes P_{r}B_{r}-\frac{3}{8}\lambda^{4}{P_{0}}^{2}P_{r}\otimes P_{r}B_{r}\\&&+\frac{3}{4}\lambda^{4}P_{0}{P_{r}}^{2}\otimes P_{0}B_{r}+\frac{3}{4}\lambda^{4}P_{0}P_{r}\otimes P_{0}P_{r}B_{r}\\&&+\frac{11}{144}\lambda^{4}P_{r}B_{r}\otimes{P_{r}}^{3}-\frac{3}{4}\lambda^{4}{P_{r}}^{2}B_{r}\otimes{P_{0}}^{2}\\&&-\frac{3}{4}\lambda^{4}{P_{r}}^{2}\otimes{P_{0}}^{2}B_{r}-\frac{5}{1152}\lambda^{4}B_{r}\otimes{P_{r}}^{4}\\&&-\frac{3}{8}\lambda^{4}P_{r}B_{r}\otimes{P_{0}}^{2}P_{r}-\frac{3}{8}\lambda^{4}P_{r}\otimes{P_{0}}^{2}P_{r}B_{r}
\end{eqnarray*}

\begin{eqnarray*}
&&\Delta P_{r}=P_{r}\otimes 1+1\otimes P_{r}+\lambda P_{r}\otimes P_{0}+\frac{1}{2}\lambda^{2}P_{r}\otimes{P_{0}}^{2}\\&&-\frac{1}{8}\lambda^{2}P{r}\otimes{P_{r}}^{2}+\frac{3}{8}\lambda^{2}{P_{r}}^{2}\otimes P_{r}+\frac{1}{4}\lambda^{3}{P_{r}}^{3}\otimes P_{0}\\&&+\frac{3}{8}\lambda^{3}P_{r}\otimes P_{0}{P_{r}}^{2}+\frac{3}{4}\lambda^{3}{P_{r}}^{2}\otimes P_{0}P_{r}+\frac{1}{2}\lambda^{4}{P_{r}}^{3}\otimes{P_{0}}^{2}\\&&-\frac{49}{1152}\lambda^{4}P_{r}\otimes{P_{r}}^{4}+\frac{11}{36}\lambda^{4}{P_{r}}^{3}\otimes{P_{r}}^{2}-\frac{1}{8}\lambda^{4}P_{r}\otimes{P_{0}}^{4}\\&&+\frac{7}{16}\lambda^{4}P_{r}\otimes{P_{0}}^{2}{P_{r}}^{2}+\frac{1}{18}\lambda^{4}{P_{r}}^{2}\otimes{P_{r}}^{3}\\&&+\frac{167}{1152}\lambda^{4}{P_{r}}^{4}\otimes P_{r}+\frac{3}{4}\lambda^{4}{P_{r}}^{2}\otimes{P_{0}}^{2}P_{r}
\end{eqnarray*}

\begin{eqnarray*}
&&\Delta P_{0}=P_{0}\otimes 1+1\otimes P_{0}+\lambda P_{r}\otimes P_{r}+\frac{1}{2}\lambda^{2}P_{0}\otimes{P_{0}}^{2}\\&&+\frac{1}{2}\lambda^{2}{P_{0}}^{2}\otimes P_{0}-\frac{1}{2}\lambda^{2}P_{0}\otimes{P_{r}}^{2}-\lambda^{2}P_{0}P_{r}\otimes P_{r}\\&&+\frac{1}{2}\lambda^{2}{P_{r}}^{2}\otimes P_{0}-\frac{1}{8}\lambda^{3}P_{r}\otimes{P_{r}}^{3}+\frac{3}{8}\lambda^{3}{P_{r}}^{3}\otimes P_{r}\\&&+\frac{1}{2}\lambda^{3}{P_{0}}^{2}P_{r}\otimes P_{r}-\lambda^{3}P_{0}{P_{r}}^{2}\otimes P_{0}+\frac{1}{4}\lambda^{4}{P_{r}}^{4}\otimes P_{0}\\&&-\frac{1}{8}\lambda^{4}P_{0}\otimes{P_{0}}^{4}-\frac{1}{8}\lambda^{4}{P_{0}}^{4}\otimes P_{0}+\frac{1}{8}\lambda^{4}P_{0}P_{r}\otimes{P_{r}}^{3}\\&&+\frac{1}{4}\lambda^{4}P_{0}\otimes{P_{0}}^{2}{P_{r}}^{2}+\frac{3}{4}\lambda^{4}{P_{0}}^{2}{P_{r}}^{2}\otimes P_{0}-\frac{7}{8}\lambda^{4}P_{0}{P_{r}}^{3}\otimes P_{r}
\end{eqnarray*}
where again we are working to quartic order in $\lambda$.
Reassuringly the coproducts ``close", {\it i.e.} they can be expressed in terms of
the generators $B_{r}, P_{r}$ and $P_{0}$, which is considered as a key consistency
criterion \cite{majRue,lukRueg,COPRODREF}
for the description of the symmetries of a noncommutative spacetime. We shall offer some comments here below on
the awkwardly lengthy expressions these coproducts have. Before we get to that
we add one more observation concerning the fact that the results we are reporting must fall within the structure
of the $\kappa$-Poincar\'e Hopf algebra. Indeed,
it has been independently established \cite{majRue,lukRueg,gacMajid,COPRODREF}
that the $\kappa$-Poincar\'e Hopf algebra describes the symmetries of the $\kappa$-Minkowski noncommutative spacetime.
The $\kappa$-Poincar\'e Hopf algebra can present itself, as far as explicit formulas are concerned, in some rather different ways,
depending on the conventions adopted. This is
because for a Hopf algebra not only linear but also non-linear redefinitions of the generators provide
admissible ``bases". This is why it is often not easy to recognize, as in the case here of interest,
that a given set of commutation relations is actually a basis of the $\kappa$-Poincar\'e Hopf algebra.
However, having established above that the Bojowald-Paily operators $B_{r}, P_{r}$ and $P_{0}$ describe
the relativistic symmetries of the $\kappa$-Minskowski spacetime, we must now infer that $B_{r}, P_{r}$ and $P_{0}$
must give a basis of $\kappa$-Poincar\'e.
The most direct way for showing that a given set of commutation rules is a basis for $\kappa$-Poincar\'e
is to show that there is nonlinear redefinition of the generators which maps them into a known basis of $\kappa$-Poincar\'e.
In order to accomplish this task we took as reference the most used basis of $\kappa$-Poincar\'e,
the so-called bicrossproduct basis, characterized by the following commutation relations ~\cite{majRue,lukRueg}:
\begin{equation}
\begin{split}
[\mathcal{P}_{0},\mathcal{P}_{r}]=0 \, ,\quad [\mathcal{B}_{r},\mathcal{P}_{0}]=iP_{r} \, ,\\
[\mathcal{B}_{r}, \mathcal{P}_{r}] = i\frac{1-e^{-2\lambda \mathcal{P}_{0}}}{2\lambda}-i\frac{\lambda}{2}\mathcal{P}^{2}_{r} \, ,
\end{split}
\end{equation}
where we introduced the notation $\mathcal{P}_{0}$, $\mathcal{P}_{r}$, $\mathcal{B}_{r}$ for the generators
of the bicrossproduct basis.

We have obtained the relationship between the Bojowald-Paily operators $B_{r}, P_{r}$ and $P_{0}$ and
bicrossproduct-basis generators  $\mathcal{P}_{0}$, $\mathcal{P}_{r}$, $\mathcal{B}_{r}$ in terms of the
function $Z(p_{r})$ which, as shown in the previous section,
must solve Eq.\eqref{condition} in order for us to have a consistent representation of $B_{r}, P_{r}$ and $P_{0}$
on the $\kappa$-Minskowski spacetime. This relationship takes the form:
\begin{equation}\label{finalg}
\begin{split}
B_{r} = \frac{ Z(\mathcal{P}_{r}e^{\lambda \mathcal{P}_{0}})\cos(\lambda Z(\mathcal{P}_{r}e^{\lambda \mathcal{P}_{0}}))}{\mathcal{P}_{r}e^{\lambda \mathcal{P}_{0}}}\mathcal{B}_{r}, \\
P_{r} = Z(\mathcal{P}_{r}e^{\lambda \mathcal{P}_{0}}),\\
P_{0} =  \frac{\sinh(\lambda \mathcal{P}_{0})}{\lambda}+\frac{\lambda}{2}\mathcal{P}^{2}_{r}e^{\lambda\mathcal{P}_{0}} \, ,
\end{split}
\end{equation}
which (since we have an explicit result for $Z(p_{r})$ to quartic order in $\lambda$) we
we can render explicit to quartic order in  $\lambda$:
\begin{equation}
\begin{split}
B_{r} = (1+\lambda^{2}\mathcal{P}^{2}_{0}-\frac{3}{8}\lambda^{2}\mathcal{P}_{r}^{2}-\frac{3}{4}\lambda^{3}\mathcal{P}_{0}\mathcal{P}_{r}^{2}+\\-\frac{5}{4}\lambda^{4}\mathcal{P}_{r}^{2}\mathcal{P}^{2}_{0}-\frac{113}{1152}\lambda^{4}\mathcal{P}^{4}_{r})\mathcal{B}_{r}
\end{split}
\end{equation}
\begin{equation}
\begin{split}
P_{r} = \mathcal{P}_{r}+\lambda \mathcal{P}_{r}\mathcal{P}_{0}+\frac{\lambda^{2}}{2}\mathcal{P}_{r}\mathcal{P}_{0}^{2}+\frac{\lambda^{2}}{8}\mathcal{P}_{r}^{3}+\frac{3}{8}\lambda^{3}\mathcal{P}_{0}\mathcal{P}_{r}^{3}+\\+\frac{\lambda^{3}}{6}\mathcal{P}_{r}\mathcal{P}_{0}^{3}+\frac{9}{16}\lambda^{4}\mathcal{P}_{0}^{2}\mathcal{P}_{r}^{3}+\frac{\lambda^{4}}{24}\mathcal{P}_{r}\mathcal{P}_{0}^{4}+\frac{55}{1152}\lambda^{4}\mathcal{P}_{r}^{5}
\end{split}
\end{equation}
\begin{equation}
\begin{split}
P_{0} = \mathcal{P}_{0}+\frac{\lambda}{2}\mathcal{P}_{r}^{2}+\frac{\lambda^{2}}{6}\mathcal{P}_{0}^{3}+\frac{\lambda^{2}}{2}\mathcal{P}_{0}\mathcal{P}_{r}^{2}+\\+\frac{\lambda^{3}}{4}\mathcal{P}_{0}^{2}\mathcal{P}_{r}^{2}+\frac{\lambda^{4}}{120}\mathcal{P}_{0}^{5}+\frac{\lambda^{4}}{12}\mathcal{P}_{0}^{3}\mathcal{P}_{r}^{2}
\end{split}
\end{equation}

The last point we want to make in this section is connected with this issue of the choice of basis
for $\kappa$-Poincar\'e, and it will take us back to the awkwardly lengthy formulas we encountered in the description
of coproducts. Different bases of $\kappa$-Poincar\'e provide an equally acceptable
mathematical characterization
of the symmetries; however, it is known that some bases provide a more intuitive description
of the associated physical properties. The main issue here concerns the relationship between
the properties of the translation generators and the properties of the energy-momentum charges:
this relationship is rather intuitive in some bases but potentially misleading in some other basis.
Here relevant is the fact that non-linear redefinitions of generators give rise to equivalent mathematical pictures,
while non-linear redefinitions of the energy-momentum charges are physically significant.
In the context of the analysis we are here reporting these concerns take shape
by looking at the form of the deformed mass Casimir obtained for the
$P_{0}$, $P_{r}$, $B_{r}$ basis of $\kappa$-Poincar\'e.
From Eq. \eqref{bjcommut}, with $[B_{r}, P_{r}] = iP_{0}$ and $[P_{0}, P_{r}] = 0$,
it follows that this deformed mass Casimir takes the form:
\begin{equation}\label{mdr}
P^{2}_{0} = \frac{2}{\lambda^{2}}\left(\lambda P_{r}\sin(\lambda P_{r}) +\cos(\lambda P_{r})-1\right) \, .
\end{equation}
Usually one is able to obtain from the form of the mass Casimir the form of the on-shell relation
by simply replacing the translation generators with the energy-momentum charges, but in this case
we conjecture that this might be inappropriate. This conjecture comes from observing that
interpreting (\ref{mdr}) as an on-shell relation for the energy-momentum charges one is led to the puzzling picture
shown in Figure 1, with the energy which is not a monotonic function of spatial momentum (and at large spatial momentum
one has that the energy decreases as spatial momentum increases).
We would argue that the implications of Figure 1 are just as awkward as the length formulas needed to
describe results. We conjecture that eventually in the study of this scenario a different basis
will be motivated, leading to a monotonic on-shellness relation and simpler formulas for coproducts.
It would of course be particularly significant if the LQG side of this scenario provided
some input on the correct choice of basis.

\begin{figure}[h!]
\centering
\includegraphics[width=3in]{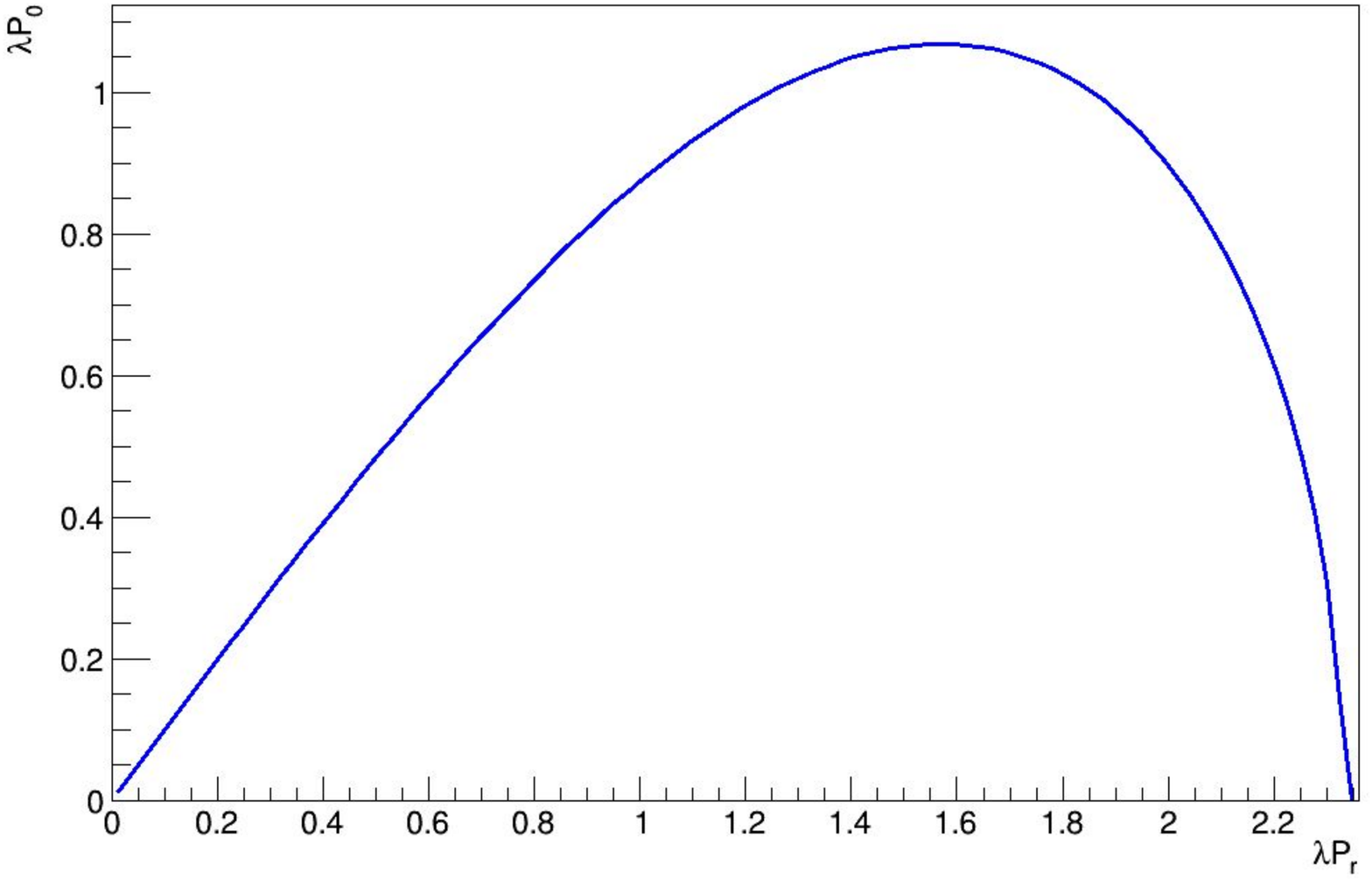}
\caption{Behavior (for $0\leq\lambda P_{r}< 2.35$)
of an on-shell relation inspired by the mass Casimir
of Eq. \eqref{mdr}.}
\label{fig}
\end{figure}

\section{OUTLOOK}
We believe the analysis here reported might be a significant step toward the description
of the Minkowski limit of LQG.
From a broader perspective we are seeing the type of back-and-forth steps that
in general will be required in order to establish a connection between top-down approaches and bottom-up
ones. The analysis reported in Ref.\cite{bojopaily}, taking LQG as starting point,
produced some commutation relations for the Minkowski limit which we could here analyze from the viewpoint
of $\kappa$-Minkowski noncommutativity. In turn, our analysis led us to establish a specific form for the coproducts, which,
as stressed above, for consistency should be found to play a role in the action of relativistic-symmetry transformations
on the product of states within the LQG formalism. We hope the challenge of seeking such a role for our coproduct
is taken by LQG experts, as it might lead to striking developments.

Similarly, through the connection with the $\kappa$-Minkowski spacetime it was natural for us to conjecture that
the translation generators adopted in Ref.\cite{bojopaily} might not be the most natural choice, at least not in the
sense that their properties reflect those of the energy-momentum charges. This was suggested by the awkwardness of
the implied on-shell relation and by the cumbersome form that the coproducts take when written in terms
of those translation generators.

Even looking beyond the contexts of LQG research and $\kappa$-Minkowski research, we have strong expectations for
 the usefulness of the strategy of analysis introduced
in Ref.\cite{bojopaily}, and here further developed. This strategy essentially sees the (modifications of the) hypersurface
deformation algebra as the point of connection between the  top-down and bottom-up
approaches. For a top-down approach obtaining results for the modifications of the
hypersurface
deformation algebra should be viewed as a very natural goal, and then, as shown here and in Ref.\cite{bojopaily}, the path
from the hypersurface
deformation algebra to a quantum-spacetime description of the Minkowski limit should be manageable. It would be particularly interesting
to see this strategy implemented in other top-down approaches besides LQG.

\section*{ACKNOWLEDGEMENTS} M. M. S. thanks the Brazilian government program ``Science without borders" for its financial support. M. R. acknowledges the support by the Accademia Nazionale dei Lincei through the Enrico Persico Prize.


\begin{thebibliography}{}

\bibitem{bojopaily} M. Bojowald, G.M. Paily, Phys. Rev. \textbf{D87}, 044044 (2013) 4.

\bibitem{gacLRR} G. Amelino-Camelia, Living Rev. Rel. \textbf{16}, (2013) 5.

\bibitem{RovelliLRR} C. Rovelli, {\em  Living Rev. Rel.} {\bf 1}, 1
            (1998).

\bibitem{AshLew} A. Ashtekar and J. Lewandowski, Class. Quant. Grav. \textbf{R53}, 21 (2004).

\bibitem{Thiem} T. Thiemann, Lect. Notes Phys. \textbf{631}, 41 (2003).

\bibitem{gacMajid} G. Amelino-Camelia, S. Majid, Int. J. Mod. Phys. \textbf{A15}, 4301 (2000).

\bibitem{majid} S. Majid, {\em Class. Quant. Grav.} {\bf 5}, 1587 (1988).

\bibitem{kowNow} J. Kowalski-Glikman, S. Nowak, Int. J. Mod. Phys. \textbf{D12}, 299 (2003).


\bibitem{girellivOriti} F. Girelli, E. R. Livine and D. Oriti, {\em  Nucl. Phys.\ B} {\bf 708}, 411
            (2005).

\bibitem{MAJIDgrb} S. Majid, W. Q. Tao, Phys. Rev.\textbf{D91}, 124028 (2015).

\bibitem{bala} A.P. Balachandran, P. Padmanabhan, JHEP \textbf{1012}, 001 (2010).

\bibitem{gacEllisMavNanSak} G. Amelino-Camelia, J.R. Ellis, N.E. Mavromatos, D.V. Nanopoulos, S. Sarkar, Nature \textbf{393}, 763 (1998).

\bibitem{gacPiran} G. Amelino-Camelia, T. Piran, Phys. Rev. \textbf{D64}, 036005 (2001).

\bibitem{FREIDEL} L. Freidel, Etera R. Livine, Phys. Rev. Lett. \textbf{96}, 221301 (2006).

\bibitem{ORITI} D. Oriti, T. Tlas, Phys. Rev. \textbf{D74}, 104021 (2006).

\bibitem{scaef} B.E. Schaefer, Phys. Rev. Lett. \textbf{82}, 4964 (1999).

\bibitem{gacfreidkowsmol} G. Amelino-Camelia, L. Freidel, J. Kowalski-Glikman, L. Smolin, Phys. Rev. \textbf{D84}, 084010 (2011).


\bibitem{gacIntJ} G. Amelino-Camelia, Int. J. Mod. Phys. \textbf{D11}, 35 (2002).

\bibitem{magsmolin} J. Magueijo, L. Smolin, Phys. Rev. Lett. \textbf{88}, 190 (2002).

\bibitem{gacModP} G. Amelino-Camelia, Mod. Phys. Lett. \textbf{A9}, 3415 (1994).

\bibitem{gacPhyLett} G. Amelino-Camelia, Phys. Lett. \textbf{B510}, 255 (2001).

\bibitem{majRue} S. Majid, H. Ruegg, Phys. Lett. \textbf{B334}, 348 (1994).

\bibitem{lukRueg} J. Lukierski, H. Ruegg, W.J. Zakrzewski, Ann. Phys. \textbf{243}, (1995) 90.

\bibitem{Dirac} P. A. M. Dirac,   Proc. Roy. Soc. Lond.\ A {\bf 268}, 57
             (1962).
\bibitem{ADM} R. L. Arnowitt, S. Deser and C. W. Misner,  Gen. Rel. Grav. {\bf 40}, 1997
             (2008).

\bibitem{CorichiReyes}  A. Corichi and J. D. Reyes, Class. Quant. Grav.   \textbf{32},  195024 (2015).

\bibitem{CaitellMielcBarr} T. Cailleteau, J. Mielczarek, A. Barrau, J. Grain, Class. Quant. Grav. \textbf{29}, 095010 (2012).

\bibitem{AshtLewMarMouThie} A. Ashtekar, J. Lewandowski, D. Marolf, J. Mourao and T. Thiemann, J. Math. Phys. {\bf 36}, 6456 (1995).

\bibitem{bojopail2} M. Bojowald, G.M. Paily, Phys. Rev. \textbf{D86}, 104018 (2012).

\bibitem{bojoHossKag} M. Bojowald, G.M. Hossain, M. Kagan, S. Shankaranarayanan, Phys. Rev. \textbf{D78}, 063547 (2008).

\bibitem{Calcagni} G. Calcagni, L. Papantonopoulos, G. Siopsis, N. Tsamis, Lect. Notes Phys. \textbf{863}, 149 (2013).

\bibitem{Barrau} A. Barrau, M. Bojowald, G. Calcagni, J. Grain, M. Kagan, JCAP \textbf{05}, 1505 (2015).

\bibitem{PerezPranz} A. Perez and D. Pranzetti,   Class. Quant. Grav {\bf 27}, 145009
             (2010).
\bibitem{Loll} R. Loll,  Class. Quant. Grav. {\bf 15}, 799
             (1998).


\bibitem{Thiemann2} T. Thiemann,  Class. Quant. Grav. {\bf 23}, 2211
             (2006).

\bibitem{CuttSakell} P. D. Cuttell, M. Sakellariadou, Phys. Rev. \textbf{D90}, 104026 (2014).


\bibitem{bojoBraRey} M. Bojowald, S. Brahma and J. D. Reyes,  Phys. Rev.\ D {\bf 92}, 045043
             (2015).

\bibitem{AlesciCianf}   E. Alesci and F. Cianfrani, Phys. Rev.  \textbf{D90}, 024006 (2014).

\bibitem{Beng} I. Bengtsson, Class. Quantum Grav. \textbf{8}, 1847 (1991).
M. Varadarajan, Class. Quantum Grav. \textbf{8}, L235 (1991).

\bibitem{BY}  D. Brown, J. W. York,  Phys. Rev. \textbf{D47}, 1407 (1993).

\bibitem{ReggeTeit} T. Regge and C. Teitelboim, {\em   Annals Phys.} {\bf 88}, 286
             (1974).

\bibitem{Mielcz} J. Mielczarek, Europhys. Lett. \textbf{108},  40003 (2014).

\bibitem{KovMeljPachol} D. Kovacevic, S. Meljanac, A. Pachol, R. Strajn, Phys. Lett. \textbf{B711}, 122 (2012).

\bibitem{COPRODREF}	A. Blaut, M. Daszkiewicz, J. Kowalski-Glikman, S. Nowak, Phys. Lett. \textbf{B582}, 82 (2004).

\end{thebibliography}
\end{document}